\documentclass[12pt]{article}
\usepackage{graphics,amssymb,float}
\usepackage{graphicx}% Include figure files
\usepackage{caption}
\usepackage{rotating}% Include figure files
\usepackage{dcolumn}% Align table columns on decimal point
\usepackage{bm}% bold math
\usepackage{cite}
\usepackage{amsmath}
\usepackage{indentfirst} % activate indent in the first paragraph
\usepackage{epstopdf}
\usepackage[usenames,dvipsnames]{xcolor}
\usepackage{setspace}
\usepackage{tabulary}
\usepackage{ulem}
\usepackage{hhline}
\usepackage{float}

\makeatletter

\@addtoreset{equation}{section}
\makeatother

\def\lsim{\;\raise0.3ex\hbox{$<$\kern-0.75em\raise-1.1ex\hbox{$\sim$}}\;}
\def\gsim{\;\raise0.3ex\hbox{$>$\kern-0.75em\raise-1.1ex\hbox{$\sim$}}\;}
\def\beq{\begin{equation}}   \def\eeq{\end{equation}}
\def\ba{\begin{array}}       \def\ea{\end{array}}
\def\bea{\begin{eqnarray}}   \def\eea{\end{eqnarray}}

\def\nl{\newline}

\begin{document}

\begin{titlepage}
\begin{flushright}
LUPM 22-024\\
BONN-TH-2022-20\\
\end{flushright}

\begin{center}
\vspace{1cm}

{\Large\bf \boldmath $M_W$, Dark Matter and $a_\mu$ in the NMSSM}

\vspace{2cm}

{\bf{Florian Domingo$^a$, Ulrich Ellwanger$^b$, Cyril Hugonie$^c$}}\\
\vspace{1cm}
\it $^a$ Bethe Center for Theoretical Physics $\&$ Physikalisches Institut der Universit\"at Bonn,
Nussallee 12, D-53115 Bonn, Germany\\
domingo@physik.uni-bonn.de\\
\it $^b$ IJCLab,  CNRS/IN2P3, University  Paris-Saclay, F-91405  Orsay,  France\\
ulrich.ellwanger@ijclab.in2p3.fr\\
\it $^c$ LUPM, UMR 5299, CNRS/IN2P3, Universit\'e de Montpellier, F-34095 Montpellier, France\\
cyril.hugonie@umontpellier.fr

\end{center}
\vspace{2cm}

\begin{abstract}

We study regions in the parameter space of the NMSSM which are able to simultaneously explain the current measured values for the $W$ mass $M_W$ and the muon anomalous magnetic moment $a_\mu$, and provide a dark matter relic density consistent with the observations  as well as constraints from detection experiments. The corresponding regions feature light charginos and sleptons in the 100~GeV$-$1~TeV range, at least some of them with masses below 175~GeV such that the electroweakly-interacting SUSY particles generate sufficiently large contributions to $M_W$. The LSP is always singlino-like with a mass below 150~GeV, and could possibly remain invisible even at future detection experiments. Decays of electroweak sparticles proceed through cascades via staus and/or sleptons which makes their detection challenging. We propose benchmark points for future searches of such sparticles.
The lightest CP-even scalar may have a mass in the 95$-$98~GeV range with, however, modest signal rates in view of the mild excesses reported in this range at LEP and by CMS at the LHC.

\end{abstract}

\end{titlepage}

\section{Introduction}

An updated measurement of the $W$~boson mass by the CDF collaboration at the Tevatron \cite{CDF} returned the result $M_W=80.4335 \pm 0.0094$~GeV, with substantially reduced uncertainties compared to earlier measurements and averages. Combining this recent CDF result with former measurements at LEP, the Tevatron and the LHC, the new world average becomes \cite{deBlas:2022hdk}
\beq\label{mwmeasured}
M_W^{\text{Exp}} = 80.4133 \pm 0.0080\ \text{GeV}\; .
\eeq
Of course, the origin of the tension between the new CDF result and the average of the former measurements from LEP, the Tevatron and the LHC remains to be analysed before any strong implications are drawn concerning the viability of models of particle physics.

On the other hand, a fit of electroweak precision data within the Standard Model (SM) gives \cite{deBlas:2021wap}
\beq\label{eq:mwsm}
M_W^{\text{SM}} = 80.3545 \pm 0.0057\ \text{GeV}\; .
\eeq
If of merely statistical nature, the difference $\Delta M_W$ between $M_W^{\text{Exp}}$ and $M_W^{\text{SM}}$ settles at $\sim 6\,\sigma$ and calls (if taken at face value) for  some contribution from physics beyond the Standard Model (BSM) to the $W$~boson mass. 

Theories beyond the SM  are more  particularly motivated if they simultaneously explain additional phenomena that are also impossible or difficult to explain within the SM. The presence of dark matter -- compatible with present constraints from direct detection experiments -- is a striking (perhaps the most striking) phenomenon of this kind. Another example is the anomalous magnetic moment of the muon $a_\mu \equiv (g_\mu-2)/2$, where the measured value (combining the BNL E821 \cite{Muong-2:2006rrc} and the Fermilab Muon g-2 \cite{Muong-2:2021ojo} experiments) deviates by $4.2\,\sigma$ from the SM.

Supersymmetric (SUSY) extensions of the SM -- in particular non-minimal versions thereof -- can address the latter two deviations from the SM phenomenology, while being in addition motivated by an alleviation of the hierarchy problem of the SM. (Existing lower bounds on masses of supersymmetric particles from fruitless searches prevent SUSY from addressing the hierarchy problem in a fully exhaustive way; some fine tuning at least in the percent range would thus persist.) 

R-parity conserving SUSY nearly automatically leads to the existence of weakly- or feebly-interacting dark matter in the form of the LSP (lightest SUSY particle), usually a mixture of binos, winos or higgsinos within the Minimal Supersymmetric extension of the SM (MSSM), or in the form of singlinos within the Non-Minimal Supersymmetric extension of the SM (NMSSM). However, in the MSSM, it is difficult to reconcile the observed relic density $\Omega h^2 = 0.1200 \pm 0.0012$ \cite{Planck:2018vyg} with upper limits from \cite{CRESST:2015txj,DarkSide:2018bpj,XENON:2018voc,XENON:2019rxp,PICO:2019vsc,PandaX-4T:2021bab} on direct detection cross sections \cite{Barman:2022jdg}
(though not impossible; see e.g.~\cite{Endo:2021zal,Chakraborti:2021squ,Ellis:2022emx}.)

On the other hand, the observed relic density and the upper limits on direct detection cross sections can be explained without contradiction in the NMSSM when considering the scenario of singlino-dominated dark matter \cite{Ellwanger:2018zxt,Cao:2018rix,Domingo:2018ykx,Abdallah:2019znp,Cao:2019qng,Wang:2020dtb,Guchait:2020wqn,KumarBarman:2020ylm,Barman:2020vzm,Zhou:2021pit,Abdughani:2021pdc,Cao:2021tuh,Wang:2022lxn,Tang:2022pxh}. 
In fact, also the computed anomalous magnetic moment $a_\mu$ of the muon can be brought near the corresponding measured value in the NMSSM \cite{Cao:2021tuh,Tang:2022pxh,Cao:2022chy,Cao:2022htd}. 
($a_\mu$ can also be explained in the MSSM, see e.g.~\cite{Athron:2021iuf,Baum:2021qzx,Chakraborti:2021kkr,Bagnaschi:2022qhb,Yang:2022gvz}, at least as long as one allows for a dark matter abundance below the observed relic density.)

Furthermore, the extended SUSY spectrum also affects the predictions for the $W$~boson mass.
Corresponding calculations have been performed in \cite{Pierce:1996zz,Heinemeyer:2002jq,Heinemeyer:2004gx,Heinemeyer:2006px,Heinemeyer:2007bw,Heinemeyer:2013dia,Bagnaschi:2022qhb,Chakraborti:2021kkr,Yang:2022gvz} and references therein for the MSSM, and for the NMSSM in \cite{Domingo:2011uf,Stal:2015zca,Athron:2022isz}. The possibility to describe the new world average for $M_W$ together with the anomalous magnetic moment of the muon and constraints on the relic density was discussed in \cite{Yang:2022gvz} for the MSSM, and in \cite{Tang:2022pxh} for the NMSSM. 

Using \texttt{FlexibleSUSY} \cite{Athron:2022isz} for the calculation of $M_W$
the authors in \cite{Tang:2022pxh} conclude that viable points in the parameter space of the NMSSM which simultaneously explain the new world average for $M_W$, $a_\mu$ and $\Omega h^2$ feature a LSP $\chi^0_1$ with a mass of $50-60$~GeV or $150-300$~GeV with a spin-independent direct detection cross section on protons $\sigma^{SI}_p$ with $\sigma^{SI}_p \gsim 10^{-11}$~pb, which can thus be tested by the future direct detection experiments XENONnT\cite{XENON:2020kmp}, LZ \cite{LZ:2015kxe} and DARWIN \cite{Schumann:2015cpa}.

In the present paper we explore the NMSSM in the light of the new value for $M_W^{\text{Exp}}$. 
We use an implementation of the calculation of $M_W$ in the code \texttt{NMSSMTools} \cite{Ellwanger:2004xm,Ellwanger:2005dv,NMSSMTools} that was described in \cite{Domingo:2011uf} (see Section~2). The dark matter relic density, direct detection cross sections and comparisons to upper limits (dominant are the ones on $\sigma^{SI}_p$ from \cite{PandaX-4T:2021bab}) are computed with the help of the program \texttt{micrOMEGAs} \cite{Belanger:2013oya}. 
The calculation of $a_\mu$ provided within \texttt{NMSSMTools} follows the description in \cite{Domingo:2008bb}.

We find new viable points in the parameter space of the NMSSM which simultaneously explain $M_W$, $a_\mu$ and $\Omega h^2$  (without obviously violating existing collider constraints), many of which
have a direct detection cross section $\sigma^{SI}_p$ below the reach of future direct detection experiments \cite{XENON:2020kmp,LZ:2015kxe,Schumann:2015cpa} and even below the neutrino floor \cite{Billard:2013qya}.
The masses of at least some of the electroweak sparticles (sleptons, staus, charginos) are in the 100-150~GeV range. Due to the dominantly singlino nature of the LSP (hence its reduced couplings to electroweakly-charged (s)particles), practically all sparticle decays proceed via the NLSP, often the tau sneutrino $\tilde{\nu}_\tau$. The corresponding cascade decays, and the small mass difference between the electroweak sparticles and the NLSP, reduce the impact of typical electroweakino (=~chargino, neutralino and slepton) searches at the LHC (or LEP) on such spectra, which we discuss in Section~3. Limits from LHC runs at 8 and 13~TeV are verified using \texttt{SModelS-2.2.0} \cite{Kraml:2013mwa,Dutta:2018ioj,Khosa:2020zar,Alguero:2021dig}.
In Section~4 we present summaries of viable points and the corresponding decay cascades for some benchmark points, and conclude in Section~5.

\section{\boldmath $\Delta M_W$ in the NMSSM}

For the particle content and the Lagrangian of the NMSSM we refer the reader to the reviews in \cite{Maniatis:2009re,Ellwanger:2009dp}. Compared to the MSSM, the NMSSM contains a 5th neutralino degree of freedom (the singlino) and extra mostly $SU(2)$-singlet-like CP-even and CP-odd scalars. The mixing angles of these singlet-like states with their MSSM-like counterparts are proportional to a dimensionless coupling $\lambda$, i.e.~these states decouple for $\lambda \to 0$. This feature is particularly relevant for the neutralino sector where the singlino, if it is the LSP, represents a good dark matter candidate: its relic density can coincide with the observed one using co-annihilation or annihilation via a resonance (e.g.~a singlet-dominated scalar) in the s-channel. On the other hand, its direct detection cross section can be very small for reduced higgsino, wino and bino components, or for destructive interferences between $SU(2)$-singlet and SM-like contributions.

In this paper, we employ the calculation of $M_W$ in the NMSSM that was described in \cite{Domingo:2011uf} and which is incorporated within \texttt{NMSSMTools-6.0.0} \cite{Ellwanger:2004xm,Ellwanger:2005dv,NMSSMTools} as the subroutine \texttt{MWNMSSM.f}. We briefly summarize the main features of this implementation below. 

First, we remind the reader that the structure of the electroweak theory correlates observables such that the fine-structure constant in the Thomson limit $\alpha$, the Fermi constant $G_{\mu}$, extracted from the muon decay $\mu\to e\nu_{\mu}\bar{\nu}_e$, and the masses $M_{W,Z}$ of the massive gauge bosons cannot be chosen independently from one another. Selecting $M_W$ as the observable that should be derived from the other three, this correlation may be expressed as:
\begin{equation}
M_W^2=\frac{M_Z^2}{2}\left\{1+\left[1-\tfrac{2\sqrt{2}\pi\alpha}{G_{\mu}M_Z^2}(1+\Delta r)\right]^{1/2}\right\}
\end{equation}
where $\Delta r$ encodes the (non-QED) radiative contributions to the muon decay, consisting of self-energy, vertex and box corrections (and itself implicitly dependent on $M_W$).

In the SM, contributions to $\Delta r$ are known up to leading four-loop order, for an estimated higher-order uncertainty reducing to a few MeV at the level of $M_W$ \cite{Awramik:2003rn}. This SM prediction is conveniently parametrized in a fit formula \cite{Awramik:2003rn,Arbuzov:2005ma}.

Beyond the SM, it is convenient to take advantage of the higher accuracy achieved in contributions to $\Delta r$ of SM-type via the splitting
\begin{equation}\label{eq:DrBSM}
\Delta r^{\text{BSM}}=\Delta r^{\text{SM}}+\Delta r^{\text{NP}}\; ,\quad \Delta r^{\text{NP}}=\Delta r^{\text{BSM}}-\Delta r^{\text{SM}}\,.
\end{equation}
Here, the first term of the right-hand side of the first equation is given by the four-loop prediction in the SM, while $\Delta r^{\text{NP}}$ is diagrammatically evaluated at a less precise order (typically one-loop), with many diagrams canceling out due to the spectrum overlap of the new-physics model with the SM (in any phenomenologically realistic BSM theory). In the (N)MSSM, $\Delta r^{\text{NP}}$ receives one-loop contributions from the modified Higgs sector and, more significantly, from the additional SUSY sector: explicit expressions are provided in e.g.~\cite{Pierce:1996zz,Cao:2008rc}. Partial two-loop order corrections are also included in \texttt{MWNMSSM.f} via the parameter $\Delta\rho = \frac{\Sigma^Z(0)}{M_Z^2}-\frac{\Sigma^W(0)}{M_W^2}$ (where $\Sigma^{Z,W}(0)$ denote the transverse parts of the $Z$ and $W$ boson self-energies at vanishing momenta, respectively,
including terms mixing SM and BSM contributions or gluonic corrections to squark loops): we refer the reader to the discussion in \cite{Domingo:2011uf} and references therein.

In fact, we specialize below on spectra with a massive squark sector (as hinted by the constraints from the LHC): from dimensional analysis, $\Delta r^{\text{NP}}\approx\tfrac{M_W^2}{16\pi^2\Lambda_{\text{NP}}^2}$, where $\Delta r^{\text{NP}}$ represents new-physics contributions  
intervening at various scales $\Lambda_{\text{NP}}$. The effect targeted by the difference between Eqs.(\ref{mwmeasured}) and (\ref{eq:mwsm}) is of order $\Delta r^{\text{NP}}\approx5\cdot10^{-3}$ and it is obvious that particles at the TeV scale (and beyond) cannot provide a significant contribution. Since contributions from the extended Higgs sectors in SUSY models are also severely constrained, the new physics responsible for the shift in $\Delta r$ needs to intervene at the electroweak scale and consists of particles interacting in a purely electroweak fashion.

In view of Eq.(\ref{eq:DrBSM}), we need to combine the uncertainty associated with $\Delta r^{\text{SM}}$, of higher-order and parametric natures, and that of the BSM evaluation, of essentially higher-order type. 
Here, for simplicity, we will combine the uncertainty on the experimental average in Eq.(\ref{mwmeasured}), that on the SM evaluation (compatible with Eq.(\ref{eq:mwsm})) and a $10\%$ fluctuation at the level of the BSM contribution in quadrature: for a BSM effect shifting the $M_W$ prediction to the central value of the experimental average, this corresponds to a global uncertainty of order $12$\,MeV for $\Delta M_W\equiv M_W^{\text{Exp}}-M_W^{\text{BSM}}$. 

At this point, we stress the necessity of a consistent control over electroweak-symmetry breaking effects in the diagrammatic calculation.
Contributions to $M_W$ should indeed vanish in the $SU(2)_L$-conserving limit. In the case of squark loops, this implies a fine cancellation between diagrams involving the components of the same $SU(2)_L$-doublet:
\begin{equation}\label{eq:sqcontr}
\Delta r^{\tilde{Q}}\sim\sum_i\big[R^2_{\tilde{t}}(i,1) f(M_{\tilde{t}_i}^2) - R^2_{\tilde{b}}(i,1) f(M_{\tilde{b}_i}^2)\big]\,,
\end{equation}
where $R_{\tilde{Q}}(i,1)$ denotes the $\tilde{Q}_L$ component contained in the mass eigenstate $\tilde{Q}_i$ with mass $M^2_{\tilde{Q}_i}$ while $f$ represents a loop function (depending on further parameters and masses). 
Then, if one employs parameters originating from two different renormalization schemes, such as pole-corrected squark masses together with $\overline{\text{DR}}$ mixing angles
(which formally introduces a mismatch of higher (two-loop) order at the level of $\Delta r$), it may produce large shifts of $M_W$ that are associated to a spurious breakdown of the electroweak symmetry at high-energy. 

One measurement of this artificial symmetry-violation is provided by the squared mass splitting $\Delta M^{2}_{LL}=\sum_i\big(R^2_{\tilde{t}}(i,1) M_{\tilde{t}_i}^2 - R^2_{\tilde{b}}(i,1) M_{\tilde{b}_i}^2\big)$: 
a departure of this quantity from its expected size at $\mathcal{O}(M_Z^2)$ -- notably for squark masses in the (multi-)TeV range -- would indeed indicate an inordinate breaking of $SU(2)_L$ 
which is bound to affect Eq.(\ref{eq:sqcontr}) as well\footnote{As the SLHA output of \texttt{NMSSMTools} provides squark masses including QCD pole corrections while mixing matrices remain $\overline{\text{DR}}$, such problems could typically occur if one were to compute $\Delta M_W$ externally from the on-shell spectrum. As it is, \texttt{MWNMSSM.f} runs internally, with access to strict $\overline{\text{DR}}$ parameters.
{On the other hand, we suspect an artifact of this type to be at the origin of the results of \cite{Tang:2022pxh} where
a comparatively heavy electroweakino spectrum seems able to explain the deviation in $M_W$.}}.
As we explained before, dimensional analysis indicates that squarks at the TeV scale cannot significantly contribute to $\Delta M_W$. 

To conclude, the shift $\Delta M_W$ can only originate from a comparatively light electroweak sector (electroweakinos $\equiv$ charginos, neutralinos, left-handed sleptons and staus) with masses in the $100-200$~GeV range, a feature previously observed in the MSSM in \cite{Heinemeyer:2013dia}.

\section{Experimental Constraints}

First of all, masses of light electroweakinos have to satisfy lower bounds from LEP \cite{LEP1,LEP2}. Lower bounds on the chargino mass depend on the decay modes of this particle: assuming $\chi^\pm_1 \to W^{(*)}+\chi^0_1$ to be dominant, constraints from LEP imply $M_{\chi^\pm_1} > 103.5$~GeV \cite{LEP1}. However, for most of the NMSSM points considered here, charginos decay through a different channel (see below) in which case we impose $M_{\chi^\pm_1} > 100$~GeV. The same constraint is imposed on slepton $\tilde{\ell}$ masses (selectrons and smuons are assumed degenerate); in the case of stau ($\tilde{\tau}$) masses we require $M_{\tilde{\tau}_1} > 93.2$~GeV \cite{LEP2}.

Constraints from searches at the LHC are verified using \texttt{SModels-2.2.0} \cite{Kraml:2013mwa,Dutta:2018ioj,Khosa:2020zar,Alguero:2021dig}. Most searches for charginos $\chi^\pm_1$ and neutralinos $\chi^0_i$ ($i \geq 2$) at the LHC assume dominant decays $\chi^\pm_1 \to W^{(*)}+\chi^0_1$ and $\chi^0_i \to Z^{(*)}/H_{SM} + \chi^0_1$. However, given the presence of light lepton-sneutrinos, tau-sneutrinos, a light mostly singlet-like Higgs state $H_1$ and the small couplings of the mostly singlet-like $\chi^0_1$ for the parameter points on which we focus, the branching fractions for $\chi^\pm_1 \to W^{(*)}+\chi^0_1$ and/or $\chi^0_i \to Z^{(*)}/H_{SM} + \chi^0_1$ are always small, resulting in limited sensitivity of the ``classical'' electroweakino searches. 

We find that the following searches for electroweakinos implemented in \texttt{SModels-2.2.0} play a dominant role in the scenario described above:

-- Searches for charginos and sleptons at 8~TeV by ATLAS in \cite{ATLAS:2014ikz,ATLAS:2014zve};

-- Searches for charginos and sleptons at 13~TeV by ATLAS in \cite{ATLAS:2018ojr,ATLAS:2018eui,ATLAS:2019lff,ATLAS:2020pgy,ATLAS:2019wgx,ATLAS:2021moa} and by CMS in \cite{CMS:2018kag,CMS:2018szt,CMS:2018eqb,CMS:2020bfa,CMS:2021edw};

-- Searches for staus by ATLAS in \cite{ATLAS:2019gti}.

Further potentially relevant searches for light electroweakinos at the LHC -- not implemented in \texttt{SModels-2.2.0} -- have been performed by CMS in \cite{CMS:2019eln,CMS:2021cox,CMS:2022sfi,CMS:2022rqk} and by ATLAS in \cite{ATLAS:2019lng,ATLAS-CONF-2022-006}. (For a recent review of searches for electroweakinos at the LHC see \cite{Adam:2021rrw}.) For a variety of reasons many of the obtained limits do not apply to the light electroweakinos considered here: assumed $\chi^0_1$ masses in \cite{CMS:2019eln,ATLAS-CONF-2022-006} are smaller, assumed chargino masses in \cite{CMS:2022sfi} are larger, assumed branching fractions for $\chi^0_i \to \chi^0_1+Z/H_{125}$ in \cite{ATLAS:2019lng,CMS:2021cox} are larger, and assumed branching fractions for $\chi^0_i \to \ell/\tau + \tilde{\ell}/\tilde{\tau}$ for $\tilde{\ell}, \tilde{\tau}$ searches in \cite{CMS:2021cox} are larger than the values obtained here after imposing the constraints implemented in \texttt{SModels-2.2.0}. Remaining relevant limits on purely left-handed stau-pairs from Fig.~7b in \cite{CMS:2022rqk} as function of $M_{\chi^0_1}$, and on purely left-handed slepton pairs from Fig.~16 in \cite{ATLAS:2019lng} and Fig.~7 in \cite{ATLAS-CONF-2022-006} are checked separately. We found that notably the constraints on purely left-handed sleptons from Fig.~7 in \cite{ATLAS-CONF-2022-006} eliminate some points in the NMSSM parameter region under discussion.
But again, limits on left-handed sleptons do not apply if these undergo dominantly cascade decays e.g.~via lighter staus or charginos which reduce the $E_\text{T}^{\text{miss}}$ of the events.

$\delta a_\mu \equiv a_\mu^{\text{Exp}}-a_\mu^{\text{SM}}$ is computed following \cite{Domingo:2008bb}. The corresponding experimental uncertainty is $\Delta a_\mu^{\text{Exp}} = \pm 0.41 \times 10^{-9}$ \cite{Muong-2:2021ojo}, while the uncertainty from its determination within the SM is $\Delta a_\mu^{\text{SM}} =\pm 0.43 \times 10^{-9}$ \cite{Aoyama:2020ynm}. We add an uncertainty from (parameter dependent) NMSSM specific contributions to $\delta a_\mu$; the combination of uncertainties amounts to about $\pm 1.55\times 10^{-9}$ at the $2\, \sigma$ level.

The presence of light electroweakinos also implies possible effects in the precise properties of SM particles, such as the $Z$- or the observed Higgs bosons. Nevertheless, the leptonic branching ratios of the $Z$-bosons measured at LEP \cite{ALEPH:2005ab} or the Higgs couplings to leptons deduced from the LHC \cite{CMS:2020xwi,ATLAS:2020fzp}
measurements come with large experimental uncertainties as compared to the magnitude of the effects discussed here, for $M_W$ or $a_{\mu}$: the impact of these observables is thus currently negligible.

Finally, we also require that viable points satisfy (at the $2\,\sigma$ level) all the phenomenological requirements implemented within \texttt{NMSSMTools}\footnote{\texttt{NMSSMTools} implements a collection of constraints from Higgs and SUSY searches (LEP, Tevatron, LHC) and low-energy observables (flavor, quarkonia, electroweak) which are listed on the website \cite{NMSSMTools}.},
in particular on the properties of the SM-like Higgs, flavor observables \cite{Domingo:2007dx,Domingo:2015wyn}, the dark matter relic density as measured by Planck \cite{Planck:2018vyg}, and constraints on direct detection cross sections using \texttt{micrOMEGAs} \cite{Belanger:2013oya}.

\section{Results and Benchmark Points}

We scanned the NMSSM parameter space using dedicated Monte Carlo programs that lead to several hundred thousand points satisfying all constraints {(at the $2\sigma$ level, as explained in the previous sections)}, only suitable subsets thereof are shown in the figures below.
The ranges of the NMSSM specific couplings, soft SUSY breaking terms and masses (with focus on the electroweak sector) are shown in Table~1. (Soft SUSY breaking terms and masses are given in GeV. For the Lagrangian of the NMSSM see \cite{Maniatis:2009re,Ellwanger:2009dp}.) 

\begin{table}[ht]
\begin{center}
\begin{tabular}{| c | c | c | c | c | }
\hline
$\lambda$& 
$\kappa$ & 
$\tan\beta$ &
$A_\lambda$ &
$A_\kappa$  \\
\hline
$0.013-0.35$  & $0.001-0.019$ & $4.5-36$ & $15-4000$ & -$196-0$\\
\hhline{|=|=|=|=|=|}
$\mu_{\text{eff}}$ &
$M_{\text{bino}} $ &
$M_{\text{wino}} $ &
$M_{\text{gluino}} $ &
$A_{\text{top}} $ \\
\hline
$118-625$ & $69-4000$ & $96-4000$ & $100-4000$ & -$4000\, - $ -1400\\
\hhline{|=|=|=|=|=|}
$A_{\tau} $ &
$A_{\mu} $ &
$ m_{\ell_L} $ &
$ m_{\tau_L}$ &
$ m_{\tau_R}$ \\
\hline
-$7-4000$ & -$2-4000$ & $89-1130$  & $83-4000$ & $83-4000$ \\
\hhline{|=|=|=|=|=|}
$m_{Q3/U3}$ &
$m_{Q2/U2} $ &
$m_{D2/D3} $ &
$m_{E2/E3} $ &
 \\
\hline
$12-4000$ & $36-4000$ & $0-4000$ & $83-4000$ & \\
\hline
\end{tabular}
\end{center}
\caption{The ranges of the NMSSM specific couplings, soft SUSY breaking terms and masses (soft SUSY breaking terms and masses in GeV) leading to points satisfying all constraints.}
\end{table}

\begin{figure}[h!]
\begin{center}
\includegraphics[scale=0.4]{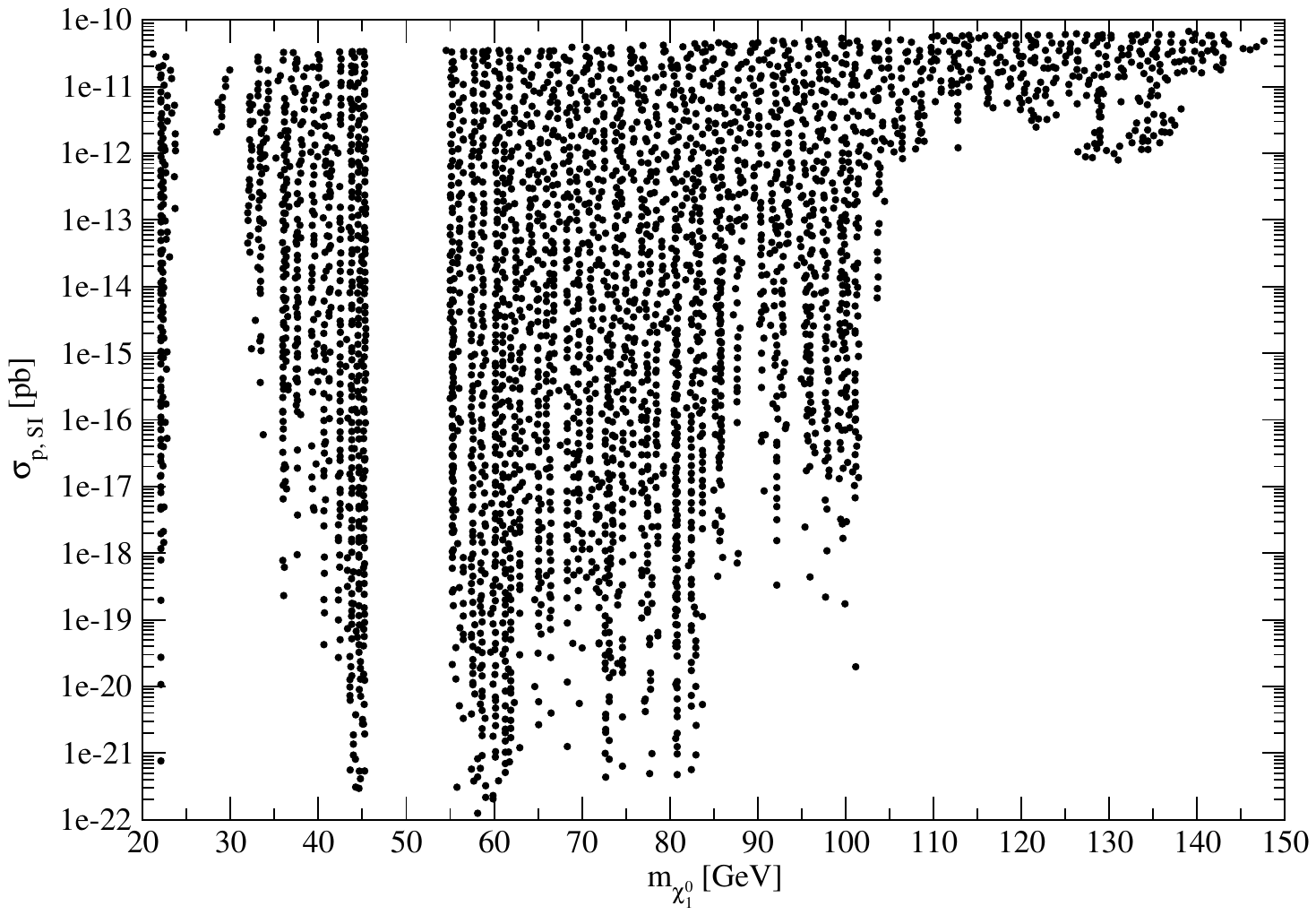}
\end{center}
\vspace*{-10mm}
\caption{Direct detection cross section $\sigma^{SI}_p$ as function of the mass of the LSP $\chi^0_1$.}
\label{fig:1}
\end{figure}

The LSP is always a mostly singlino-like neutralino with a mass {$\sim 20-150$~GeV}, and a singlino component $> 0.98$. Consequently its spin-independent direct detection cross sections on protons $\sigma^{SI}_p$ for dark matter experiments are small, possibly below $2.5\times 10^{-13}$~pb, i.e.~below the sensitivities of the future experiments XENONnT\cite{XENON:2020kmp}, LZ \cite{LZ:2015kxe} and DARWIN \cite{Schumann:2015cpa} as well as the neutrino floor. Dark matter annihilation dominantly proceeds via $Z$ or the SM Higgs boson in the s-channel, or via sneutrino co-annihilation. These cases are clearly identifiable in Fig.~\ref{fig:1} showing $\sigma^{SI}_p$ as function of the mass of the LSP $\chi^0_1$, where $M_{\chi^0_1}$ equals about half of the mass of an s-channel resonance, or more in the case of 
co-annihilation.\footnote{We note in passing that the dark matter phenomenology significantly contrasts with that identified in \cite{Tang:2022pxh}. 
We believe that the scanning procedure is at the origin of these divergent results, our scanning algorithm allowing for a better coverage of the NMSSM parameter space.}

\begin{figure}[h!]
%\vspace*{-5mm}
\begin{tabulary}{\linewidth}{CC}
\includegraphics[height=0.38\textheight]{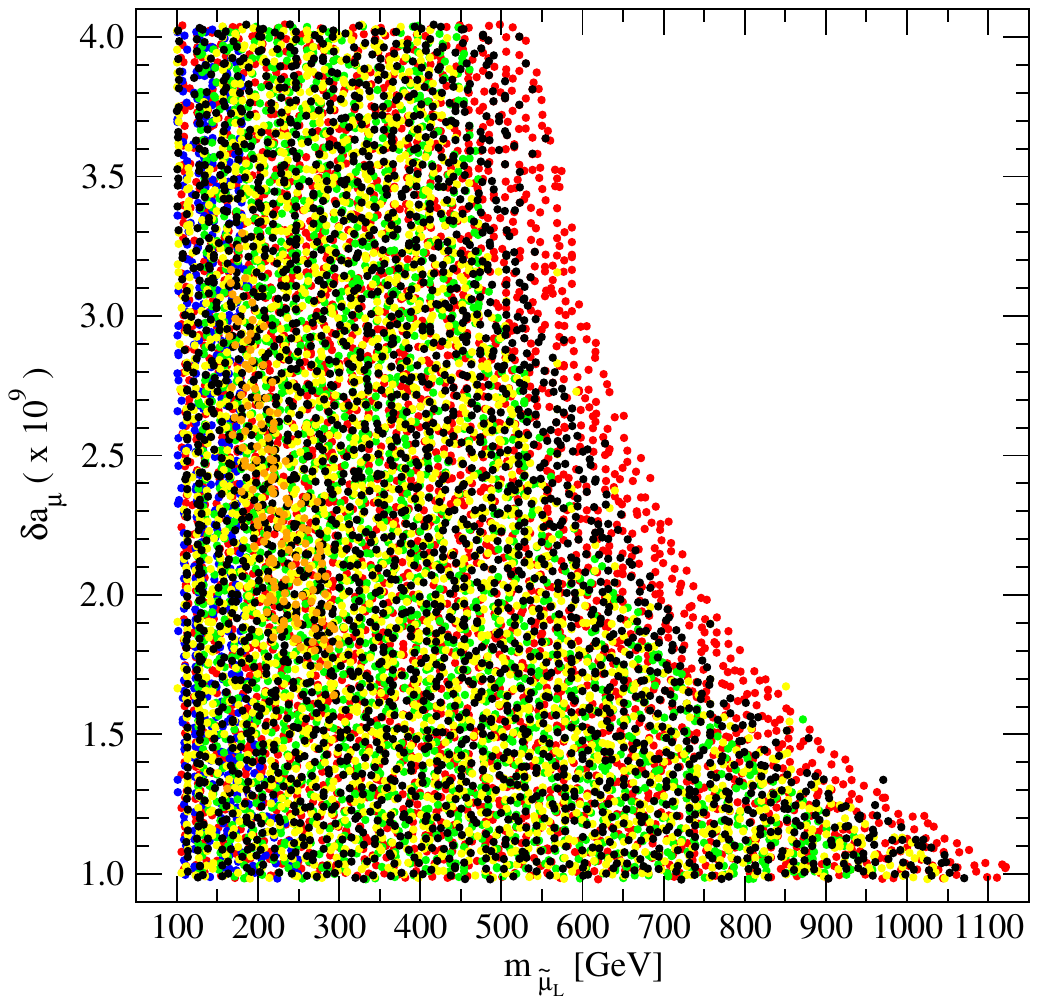}
&
\includegraphics[height=0.38\textheight]{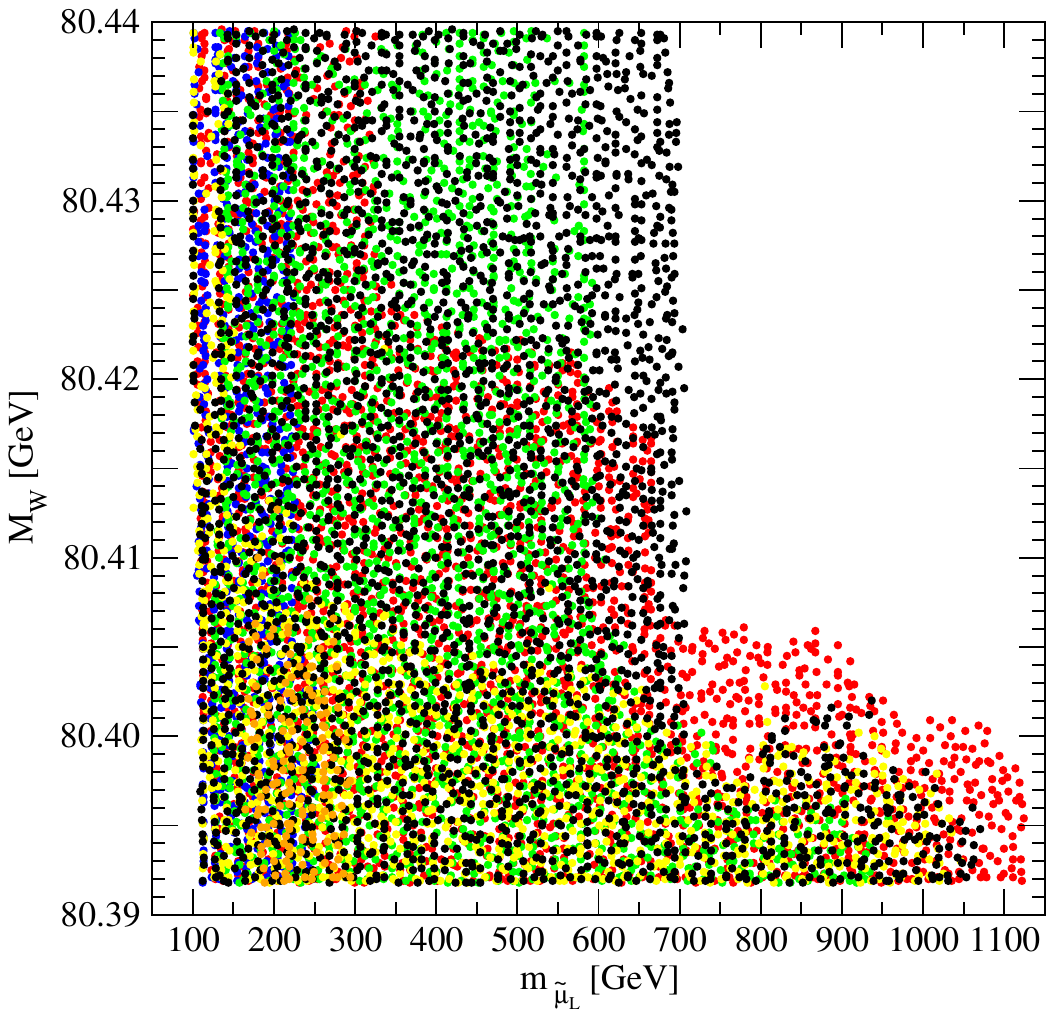}
\\ \vspace{-10mm}  Fig.~2a & \vspace{-10mm}  Fig.~2b
\end{tabulary}
\vspace*{-10mm}
\caption{$\delta a_\mu$ (left) and $M_W$ (right) as function of the left-handed smuon mass $m_{\tilde{\mu}_L}$.
The colors indicate the dominant decay modes of the chargino $\chi^\pm_1$: blue for $\ell + \tilde{\nu}_\ell$, red for $\tau + \tilde{\nu}_\tau$, green for $\chi^0_1 + W^{(*)}${, yellow for $\chi^0_2 + W^{(*)}$, orange for $\nu_\ell+\tilde{\ell}$} and black for $\nu_\tau+\tilde{\tau}$.}
\label{fig:2}
\end{figure}

In Figs.~2 we show $\delta a_\mu \equiv a_\mu^{\text{Exp}}-a_\mu^{\text{SM}}$ and $M_W$ as function of the left-handed smuon mass $m_{\tilde{\mu}_L}$ for viable points. (Recall that we assume degenerate selectrons and smuons.) Fig.~2a is consistent with the idea that a lower limit on $\delta a_\mu$ (depending on the combined experimental and theoretical uncertainties) results in an upper bound on $m_{\tilde{\mu}_L}$. Similarly, from Fig.~2b one infers that only a $2\,\sigma$ uncertainty on $M_W$ larger than {7}~MeV allows for $m_{\tilde{\mu}_L}>$ {700}~GeV.

Points in the parameter space of the NMSSM which satisfy all the above constraints still differ in the dominant decay modes of the lightest chargino $\chi^\pm_1$ among which we count {$\ell + \tilde{\nu}_\ell$, $\tau + \tilde{\nu}_\tau$, $\chi^0_{1,2} + W^{(*)}$, $\nu_\tau+\tilde{\tau}$ and only occasionally $\nu_\ell+\tilde{\ell}$}.
In view of future investigations at the LHC (addressing searches of the chargino or of heavier particles decaying through it), the identification of the dominant decay modes of $\chi^\pm_1$ will play an important role. Therefore we specify these decay modes with distinct colors in the Figures 2-3: blue for $\ell + \tilde{\nu}_\ell$, red for $\tau + \tilde{\nu}_\tau$, green for $\chi^0_1 + W^{(*)}${, yellow for $\chi^0_2 + W^{(*)}$, orange for $\nu_\ell+\tilde{\ell}$} and black for $\nu_\tau+\tilde{\tau}$.

In Figs.~3 we show the masses of the chargino $\chi^\pm_1$ and the left-handed stau $\tilde{\tau}_L$ as function of the left-handed smuon mass for viable points. We see that $\chi^\pm_1$ and smuon masses, as well as
$\tilde{\tau}_L$ and smuon masses, cannot be simultaneously be large ({$\gsim 300$~GeV and $\gsim 630$~GeV}, respectively). The reason is that at least some electroweakinos must be light in order to generate sufficiently large radiative corrections to $M_W$. 
It is remarkable that the possible existence of such comparatively light electroweakly-interacting exotic particles is compatible with the current collider limits.

\begin{figure}[h!]
%\vspace*{-5mm}
\begin{center}
\begin{tabulary}{\linewidth}{CC}
\includegraphics[height=0.38\textheight]{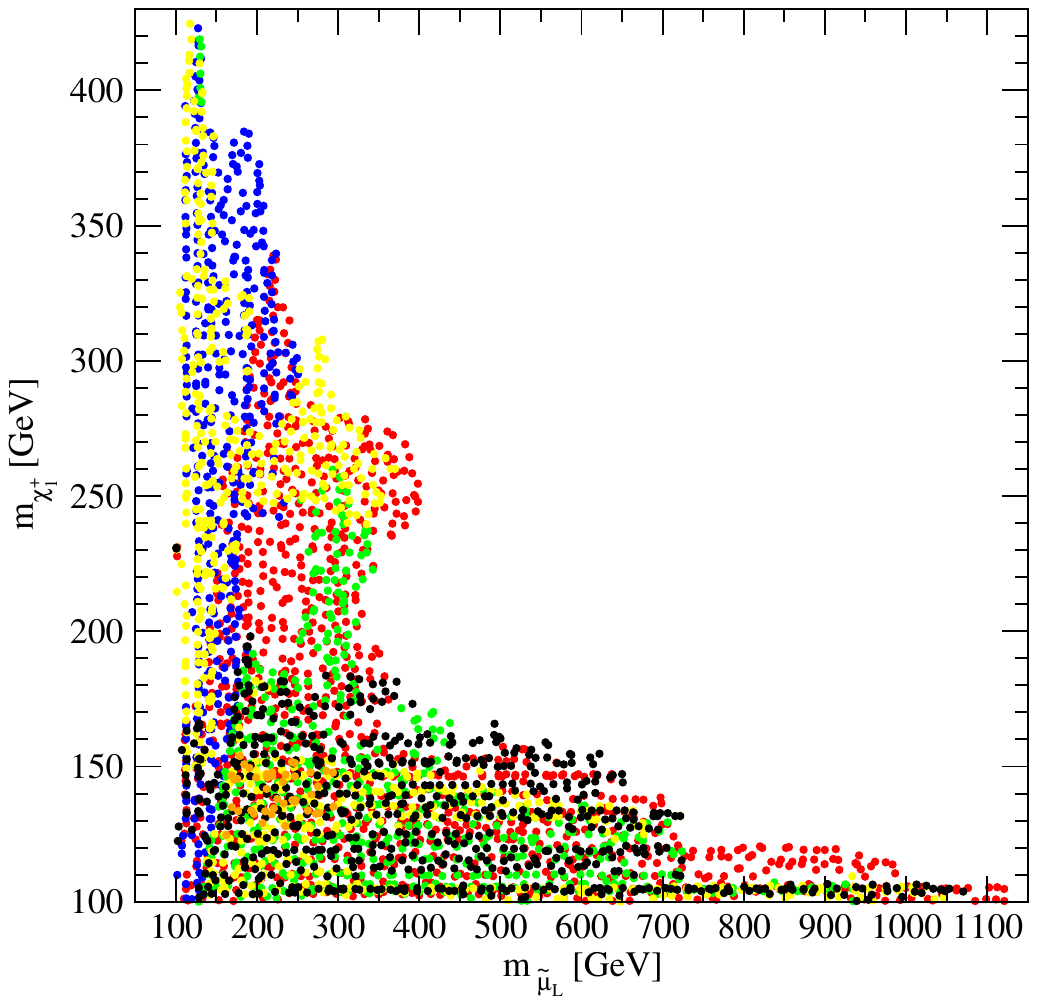}
&
\includegraphics[height=0.38\textheight]{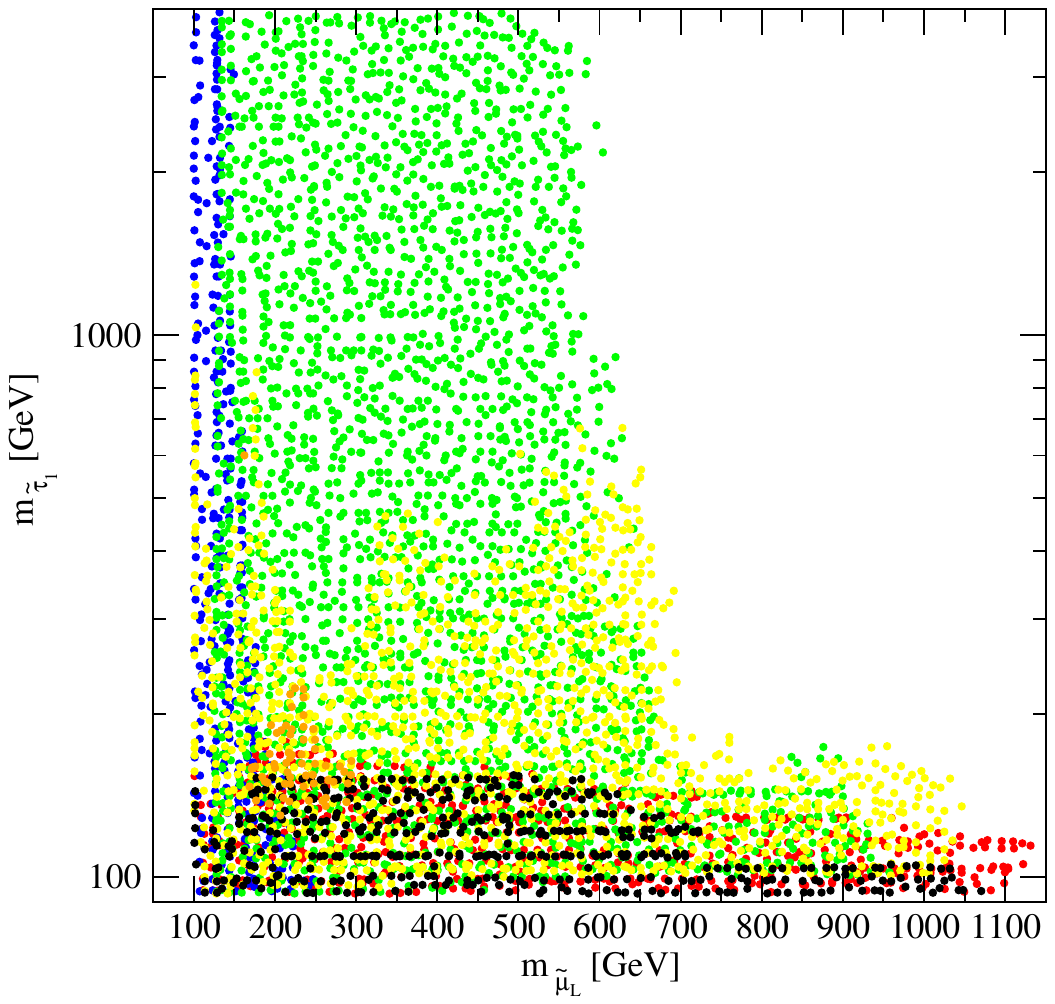}
\\ \vspace{-10mm}  Fig.~3a & \vspace{-10mm}  Fig.~3b
\end{tabulary}
\end{center}
\vspace*{-10mm}
\caption{Masses of the chargino $\chi^\pm_1$ (Fig.~3a) and the left-handed stau $\tilde{\tau}_L$ (Fig.~3b) as function of the left-handed smuon mass.
The color code is as in Fig.~\ref{fig:2}.}
\label{fig:3}
\end{figure}

The viable points in the parameter space of the NMSSM obtained here always feature a mostly singlet-like Higgs scalar $H_1$ with a mass below 125~GeV. It is checked in \texttt{NMSSMTools} that its direct production rate satisfies constraints from LEP and the LHC, and that the branching fractions of the SM Higgs to $H_1 + H_1$ satisfy constraints from ATLAS and CMS. In Fig.~4 we show the magnitude of the singlet component squared $S_{13}^2$ as function of its mass and observe that, for the viable points, $S_{13}^2$ is always above 0.93, i.e.~the doublet component squared (needed in its direct production of this state in any process at LEP and the LHC) is always below 0.07, implying a challenging detectability at colliders.
\begin{figure}[h!]
%\vspace*{-5mm}
\begin{center}
\includegraphics[scale=0.4]{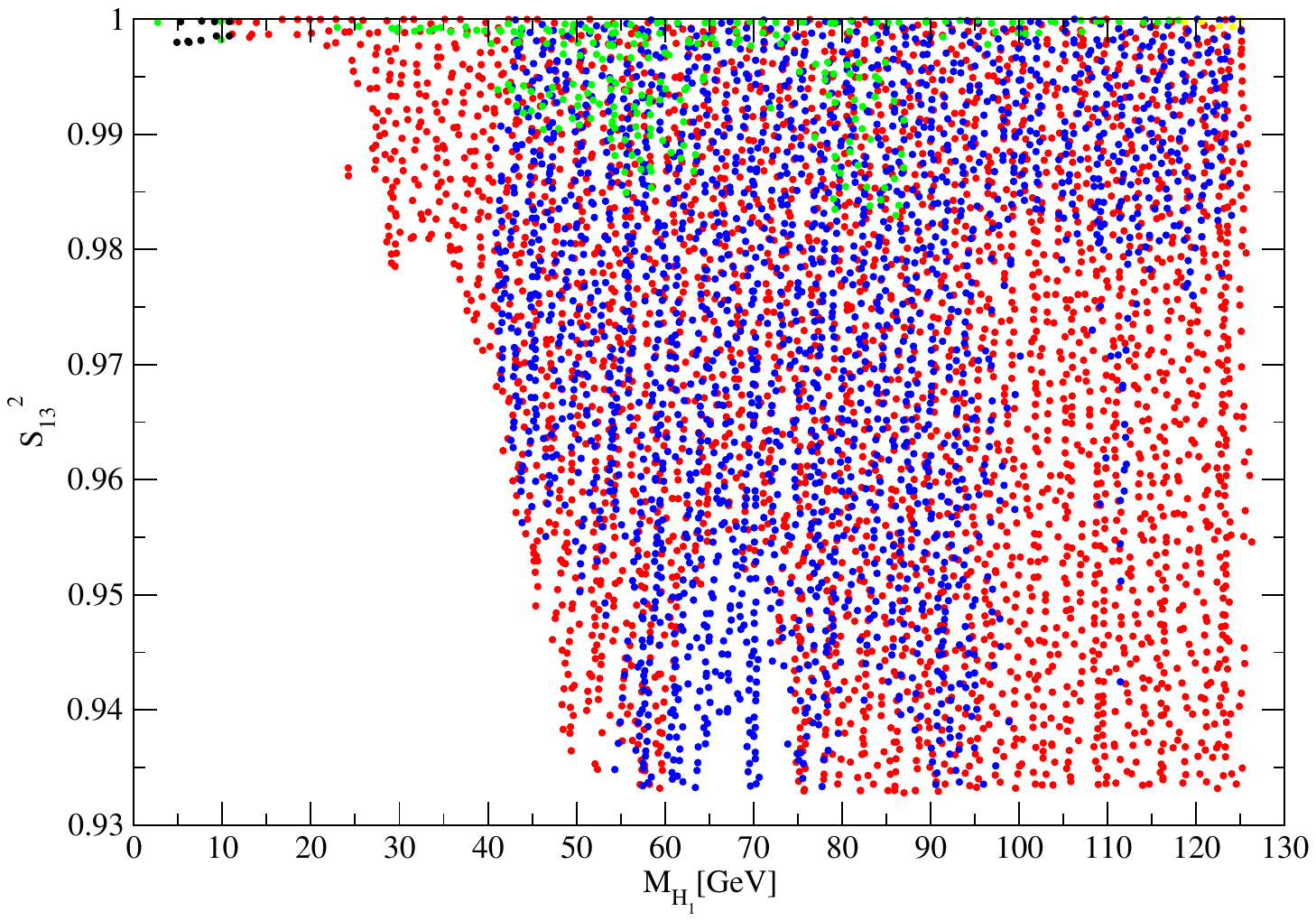}
\end{center}
\vspace*{-10mm}
\caption{$S_{13}^2$ as function of $M_{H_1}$. The colors indicate the dominant decay modes of $H_1$: $b\bar{b}$ for red points, $A_1A_1$ for blue points, 2 jets (gluon gluon or $c\bar{c}$) for green points, {$WW^*$ for yellow points and} $\tau\bar{\tau}$ for black points (for $M_{H_1}$ below the $b\bar{b}$ threshold).}
\label{fig:4}
\end{figure}

Interestingly enough, the mass of $H_1$ possibly falls in the $95-98$~GeV range. A mild excess in the $b\bar{b}$ channel in this mass range was observed at LEP \cite{LEPWorkingGroupforHiggsbosonsearches:2003ing,ALEPH:2006tnd}, corresponding to a signal strength of $\sim 10\%$ in units of the signal strength of a SM Higgs in the same channel. Moreover, the CMS collaboration has observed a mild excess in the $\gamma \gamma$ channel in this mass range both at 8 and 13~TeV \cite{CMS:2018cyk}. Locally, the excess corresponds to a signal strength of $\sim 0.5 \pm 0.18$ in units of the signal strength of a SM-like Higgs boson in the same channel. Within the NMSSM, the excess at LEP was first discussed in \cite{Belanger:2012tt}. A possible description of the excess at CMS within the NMSSM was proposed in \cite{Cao:2016uwt,Wang:2018vxp,Choi:2019yrv,Cao:2019ofo,Hollik:2020plc,Biekotter:2021qbc}. Clearly, the production of a mostly singlet-like $H_1$ at LEP or the LHC requires some singlet-doublet mixing. Such a mixing involving the SM-like Higgs boson at 125~GeV reduces its couplings $\kappa_{W,Z}$ to the electroweak gauge bosons. Such a reduction is severely constrained by the recent measurements of properties of the SM Higgs boson by ATLAS \cite{ATLAS:2022vkf} and CMS \cite{CMS:2022dwd} which, taken together, imply $\kappa_{W,Z}> 0.966$ at the $2\, \sigma$ level. Applying this constraint, we find that $H_1$ in the $95-98$~GeV range can have a reduced signal strength of $\sim 6\%$ in the $b\bar{b}$ channel at LEP and, simultaneously, a reduced signal strength of $\sim 8\%$ in the $\gamma \gamma$ channel at the LHC. (The constraint on the $H_1 - H_{125}$ mixing angle also explains why $S_{13}^2$ is bounded from below by $\sim 0.93$ in Fig.~4.) On the other hand, we could not discover points leading to a sizable signal in the $\tau^+\tau^-$ channel, as motivated in the same mass range by the searches from CMS \cite{CMS:2022goy}.

Stop decay cascades are initiated by decays into either $b$~quarks + charginos or top quarks + neutralinos $\chi^0_{2,3}$, but charginos and neutralinos $\chi^0_{2,3}$ often undergo lepton-rich or tau-rich subsequent decays.
(Cascade decays via the stau sector have also been found in \cite{Yang:2022gvz} for viable points in the MSSM where, however, the relic density is below the observed one.)
The extra MSSM-like neutral and charged Higgs bosons also have masses in the $1-4$~TeV range.

In order to illustrate the various decay modes in the electroweak sector we show the masses and branching fractions for four benchmark points BP1-BP4 in the Tables below. These points differ in the dominant decay modes of the lighter chargino $\chi^\pm_{1}$. (BP2 and BP4 are invisible even in future dark matter detection experiments \cite{XENON:2020kmp,LZ:2015kxe,Schumann:2015cpa}.)

\begin{table}[ht]
\begin{center}
\begin{tabular}{| c | c | c | c | c | c | c | c | c | c |  }
\hline
&$m_{\chi^\pm_1}$  & $m_{\chi^\pm_2}$  &
$m_{\chi^0_1}$ &$m_{\chi^0_2}$ &$m_{\chi^0_3}$ &
$m_{\tilde{\ell}_L}$ & $m_{\tilde{\nu}_\ell}$ &
$m_{\tilde{\tau}_1}$ &$m_{\tilde{\nu}_\tau}$  
\\
\hline
BP1 & 
107 & 307 & 
58 & 95 & 114 & 
147 & 125 & 
97 & 105  \\
\hline
BP2 & 
139 & 949  & 
58 & 75 & 147 &
113 & 83 & 
100 & 95  \\
\hline
BP3 & 
109 & 258 & 
44 & 107 & 167 & 
140 & 116 & 
210 & 196  \\
\hline
BP4 & 
114 & 387 & 
44 & 84 & 116 & 
176 & 158 & 
110 & 79  \\
\hhline{|=|=|=|=|=|=|=|=|=|=|}
& $M_{H_1}$  & $M_{A_1}$  & $\lambda$  & $\kappa$  &
$A_\lambda$ &$A_\kappa$ & $\sigma^{SI}_p$ &
$\delta a_\mu$ & $M_W$  
\\
\hline
BP1 & 
54 & 11 & 0.089 & 0.0095 & 
2526 & -0.6 & $1.87\times 10^{-12}$ & 
$2.98\times 10^{-9}$ & 80.402 
 \\
\hline
BP2 & 
53 & 27  & 0.036 & 0.0075 & 
3310 & -8.5 & $2.18\times 10^{-15}$ & 
$2.04\times 10^{-9}$ & 80.406 
\\
\hline
BP3 & 
41 & 29 & 0.123 & 0.013 & 
2245 & -10.5 & $2.77\times 10^{-11}$ & 
$2.99\times 10^{-9}$ & 80.404
\\
\hline
BP4 & 
47 & 56 & 0.211 & 0.013 & 
387 & -42.5 & $3.90\times 10^{-16}$ & 
$2.40\times 10^{-9}$ & 80.401
\\
\hline
\end{tabular}
\end{center}
\caption{Sparticle and mostly singlet-like Higgs boson masses (in~GeV), NMSSM specific parameters,
the spin-independent direct detection cross section of the LSP on protons (in pb), $\delta a_\mu$ and $M_W$ for four benchmark points. Leptons $\ell$ denote both electrons and muons.}
\end{table}

BP1 corresponds to a case where $\chi^+_1$ and $\chi^0_2$ have sizeable wino components 0.9 and 0.5, respectively. The dominant decay of $\chi^+_1$ is $\chi^+_1 \to  \nu_\tau+\tilde{\tau}^+$ with a branching ratio of $72\%$, $\tilde{\tau}^+$ decays to $\tau^+ +\chi^0_1$ with a branching ratio of $100\%$. $\chi^0_2$ decays dominantly into 3-body channels $\nu_\tau+\bar{\nu}_\tau+\chi^0_1$ (to $41\%$) and $\tau^+ + \tau^- +\chi^0_1$ (to $39\%$).
Hence BP1 is difficult to detect in standard search channels for charginos/neutralinos. Given the $\tilde{\tau}_1$ and $\chi^0_1$ masses shown in Table~2, the dedicated searches for staus by ATLAS in \cite{ATLAS:2019gti} and CMS in \cite{CMS:2019eln,CMS:2022rqk} are also not sensitive to BP1.

In the case of BP2, the mostly higgsino-like $\chi^+_1$ 
decays into sleptons, the tau sneutrino and stau with branching fractions shown in Table~3. The same holds for the higgsino-like neutralinos $\chi^0_{3,4}$ which are dominantly produced in association with $\chi^+_1$. Sleptons, tau sneutrino and stau decay into the bino-like neutralino $\chi^0_2$ which finally decays into $\chi^0_1+W^*$. Such lengthy cascade decays into lepton/tau-rich final states are challenging to detect. Interestingly, mild (non-significant) excesses have been observed in search channels for $2-3$ leptons plus $E_T^{\text{miss}}$ by ATLAS in \cite{ATLAS:2019lng,ATLAS:2021moa} and by CMS in \cite{CMS:2021edw}.
\begin{table}[ht]
\begin{center}
\begin{tabular}{| c | c | c | c | }
\hline
$BR(\chi^+_1 \to \ell^+ +\tilde{\nu}_\ell)$& 
$BR(\chi^+_1 \to \tau^+ +\tilde{\nu}_\tau)$ & 
$BR(\chi^+_1 \to  \nu_\tau+\tilde{\tau}^+)$ &
$BR(\chi^0_3 \to  \tau^+ +\tilde{\tau}^-)$    \\
\hline
 0.172  & 0.419 & 0.212 & 0.149\\
\hhline{|=|=|=|=|}
$BR(\chi^0_3 \to  \nu_\ell+\tilde{\nu}_\ell)$ &
$BR(\chi^0_4 \to  \tau^+ +\tilde{\tau}^-)$ &
$BR(\chi^0_4 \to  \ell^+ +\tilde{\ell}^-)$  &
$BR(\chi^0_4 \to  \nu_\ell +\tilde{\nu}_\ell)$ \\
\hline
0.203 & 0.201 & 0.058 & 0.094\\
\hhline{|=|=|=|=|}
$BR(\tilde{\ell} \to  \ell +\chi^0_2)$ &
$BR(\tilde{\tau} \to  \tau +\chi^0_2)$ &
$BR(\tilde{\nu}_\ell \to  \nu_\ell +\chi^0_2)$ &
$BR(\tilde{\nu}_\tau \to  \nu_\tau +\chi^0_2)$ \\
\hline
0.95 & 0.98 & 0.88  & 0.95 \\
\hline
\end{tabular}
\end{center}
\caption{Branching fractions for the relevant electroweak sparticles of BP2. $\tilde{\ell}$ denote left-handed selectrons {\it or} smuons. The neutralino $\chi^0_2$ decays to $100\%$ to $W^* +\chi^0_1$.}
\end{table}

In the case of BP3 the mostly wino-like chargino $\chi^+_1$ decays into $\chi^0_1+W^*$ as it is assumed in most searches. However, the mostly wino-like neutralino $\chi^0_2$ decays to $\chi^0_1+H_1$ with a branching fraction of $97\%$. $H_1$ with its mass of 41~GeV decays to $b\bar{b}$ with a branching fraction of $91\%$.

BP4 corresponds to a mostly wino-like chargino $\chi^+_1$ decaying with a branching fraction of $98\%$ to $\tau + \tilde{\nu}_\tau$. $\chi^0_2$ is bino-like, and the wino-like $\chi^0_3$ decays invisibly to $\nu_\tau + \tilde{\nu}_\tau$ since stau sneutrinos decay to $\nu_\tau + \chi^0_1$.

The properties of these benchmark points clarify why many scenarios in the NMSSM with light electroweakinos, but a singlino-like LSP, would have remained undetected at the LHC so far. It is of course desirable that dedicated searches investigate these unconstrained (and motivated) regions in parameter space.

\section{Conclusions and Outlook}

In the present paper we have shown which sparticle spectra in the NMSSM can simultaneously describe $\Delta M_W$, dark matter and $a_\mu$. 
They are characterized by light staus, charginos and sleptons. It is interesting that light charginos and sleptons, required for a SUSY explanation of $a_\mu$, can also lead to sizable contributions to $\Delta M_W$. Still, additional contributions from light staus are helpful if one requires that $\Delta M_W$ satisfies eq.~\eqref{mwmeasured}. 

In principle, these features could also be fulfilled in the MSSM. However, only the presence of the extra singlino-like LSP in the NMSSM and tau- or lepton-sneutrinos as NLSPs makes this scenario phenomenologically viable given the present constraints from the LHC (and Dark Matter): The couplings of all MSSM-like sparticles to the singlino-like LSP are so small that they prefer to decay via the NLSP (staus, sleptons or sneutrinos) leading to extra steps in the sparticle decay cascades and tau- or lepton-rich final states with reduced $E_T^{\text{miss}}$. These require dedicated searches and limits from more standard search channels for sparticles are alleviated. 

Moreover the extra singlino-like LSP in the NMSSM is a dark matter candidate allowing to describe the dark matter relic density without conflict with constraints from present, and possibly even from future direct detection experiments.
In fact, $\sigma^{SI}_p$ may fall below the neutrino floor. Hence future tests of these scenarios can and have to rely on searches for electroweakly interacting (s)particles in the $100-200$~GeV range.

\section*{Acknowledgements}

We like to thank Tian-Peng Tang and Yue-Lin Sming Tsai for discussions and information on their paper {\cite{Tang:2022pxh}}.
Our numerical results have been obtained thanks to the cloud computing infrastructure at LUPM, founded by OCEVU Labex, and France-Grilles. 
We thank Sabine Kraml, Andre Lessa and Wolfgang Waltenberger for help with \texttt{SModelS}.

\end{document}